\documentclass[preprint,aps, nofootinbib]{revtex4}
\input{epsf.tex}
\usepackage[dvips]{graphics,epsfig}
\def\be{\begin{equation}}
\def\ee{\end{equation}}
\def\ba{\begin{eqnarray}}
\def\ea{\end{eqnarray}}
\def\Box{\mathord{\dalemb{7.9}{8}\hbox{\hskip1pt}}}
\def\dalemb#1#2{{\vbox{\hrule height.#2pt
        \hbox{\vrule width.#2pt height#1pt \kern#1pt \vrule width.#2pt}
        \hrule height.#2pt}}}

\def\Box{\mathord{\dalemb{7.9}{8}\hbox{\hskip1pt}}}
\def\dalemb#1#2{{\vbox{\hrule height.#2pt
        \hbox{\vrule width.#2pt height#1pt \kern#1pt \vrule width.#2pt}
        \hrule height.#2pt}}}

\def\ba{\begin{eqnarray}}
\def\ea{\end{eqnarray}}
\def\be{\begin{equation}}
\def\ee{\end{equation}}
\def\tr{{\rm tr}}
\def\gtorder{\mathrel{\raise.3ex\hbox{$>$}\mkern-14mu
             \lower0.6ex\hbox{$\sim$}}}
\def\ltorder{\mathrel{\raise.3ex\hbox{$<$}\mkern-14mu
             \lower0.6ex\hbox{$\sim$}}}

\def\to{\rightarrow}

\def\be{\beta}

\def\frac#1#2{{\textstyle{{#1}\over {#2}}}}

\def\lsim{\mathrel{\rlap{\lower4pt\hbox{\hskip1pt$\sim$}}
    \raise1pt\hbox{$<$}}}
\def\gsim{\mathrel{\rlap{\lower4pt\hbox{\hskip1pt$\sim$}}
    \raise1pt\hbox{$>$}}}
\def\sqr#1#2{{\vcenter{\vbox{\hrule height.#2pt
         \hbox{\vrule width.#2pt height#1pt \kern#1pt
         \vrule width.#2pt}
         \hrule height.#2pt}}}}

\begin{document}

\def\Box{\mathord{\dalemb{7.9}{8}\hbox{\hskip1pt}}}
\def\dalemb#1#2{{\vbox{\hrule height.#2pt
        \hbox{\vrule width.#2pt height#1pt \kern#1pt \vrule width.#2pt}
        \hrule height.#2pt}}}

\def\ba{\begin{eqnarray}}
\def\ea{\end{eqnarray}}
\def\be{\begin{equation}}
\def\ee{\end{equation}}
\def\tr{{\rm tr}}
\def\gtorder{\mathrel{\raise.3ex\hbox{$>$}\mkern-14mu
             \lower0.6ex\hbox{$\sim$}}}
\def\ltorder{\mathrel{\raise.3ex\hbox{$<$}\mkern-14mu
             \lower0.6ex\hbox{$\sim$}}}

\rightline{DF/IST-1.2007}

\title{Spontaneous Symmetry Breaking in the Bulk 
as a Localization Mechanism of Fields on the Brane}

\author{Orfeu Bertolami}
\altaffiliation{Also at Centro de F\'\i sica dos Plasmas, Instituto Superior T\'ecnico; Email address: orfeu@cosmos.ist.utl.pt}

\author{Carla Carvalho}
\altaffiliation{Also at Centro de F\'\i sica dos Plasmas, Instituto Superior T\'ecnico; Email address: ccarvalho@ist.edu}

\vskip 0.5cm

\affiliation{Departamento de F\'\i sica, Instituto Superior T\'ecnico \\
Avenida Rovisco Pais 1, 1049-001 Lisboa, Portugal}

\vskip 1.0cm

\begin{abstract}

{We consider real and complex scalar fields 
non-minimally coupled to gravity
in the bulk spacetime
and study their impact on the brane upon acquiring a non-vanishing 
vacuum expectation value. 
When examining the case of the complex scalar 
field, a minimally coupled $U(1)$ 
gauge field is also considered 
so that spontaneous symmetry breaking can take 
place. 
Particular attention is paid to the terms arising from the 
junction conditions,
which act towards reinforcing the mechanism of 
spontaneous symmetry breaking. 
We find that the embedding of a braneworld universe in a bulk
spacetime endowed with matter fields can induce this mechanism at very high
energies, which implies the localization on the brane of the bulk fields.

}
\end{abstract}

\maketitle

\section{Introduction}

The breaking of geometrical symmetries or internal symmetries, such as
those associated with gauge invariance,   
is a crucial ingredient in physics. 
This breaking plays a central role both in condensed matter physics and in
quantum field theory. In particular, the phenomenological 
consistency of the electroweak unification relies fundamentally on the 
breaking of a gauge symmetry. It is well known that some important 
features of the symmetry breaking mechanism are altered once the coupling 
to gravity is considered. For the case of scalar fields this occurs via 
a non-minimal coupling with the scalar curvature. 
Such coupling may lead to important physical
consequences, as it affects the positivity of the Einstein tensor 
and the stability of theories coupled to gravity \cite{B87}. It is
worth stressing that the positivity  
of the Einstein tensor is a condition of the well-known singularity
theorems for gravity.  
A non-minimal coupling plays also a crucial role in some 
alternative gravity models, commonly referred 
to as induced gravity \cite{Fujii71}, as it is through this coupling
that the mechanism of symmetry breaking can generate a gravitational
coupling and an induced cosmological constant.



In this work we shall consider such non-minimally coupled scalar field models 
in the context of braneworlds. 
The two bulk fields, namely gravity and the scalar field, must be
allowed to interact in the five-dimensional spacetime and not solely on
the boundary. The simplest such interaction couples the scalar
curvature to the scalar field. 
Notice that in string theory, for instance, the couplings to the
dilaton at the affective action level involve instead an overall  
exponential function in the s-parametrization and derivative terms
\cite{SEffA}.  

Bulk scalar fields have been previously discussed in various studies
of non-vanishing vacuum bulk field configurations,
in both one \cite{DKS00, KT05} and two-brane setups \cite{GW99, chang00}.
Here we study real and complex scalar fields in the bulk spacetime 
and examine their implication for the mechanism of spontaneous symmetry
breaking (SSB) on the brane upon acquiring a non-vanishing bulk 
vacuum expectation value (vev). When studying the case of the complex scalar 
field, we also consider a minimally coupled $U(1)$ 
gauge field so that spontaneous breaking of the gauge symmetry can take 
place.
New terms arising from the junction conditions will contribute
to the equations induced on the brane and entail consequences in the
induced SSB mechanism.
Since the masses induced by the brane mediated SSB mechanism are of
order the four-dimensional Planck mass, the range of the induced
interactions become short about the brane, suggesting the localization of
the bulk fields on the brane.

This paper consists also of a preliminary study of the possibility of 
using spontaneous symmetry breaking as a means for localizing
matter on the brane. In the Goldstone mechanism, a mass is attained
in the direction along which the system undergoes spontaneous symmetry
breaking. The consequent gain in inertia would be expected to 
prevent the motion of matter along that direction and thus to constrain
matter to stay on the orthogonal directions, which would define the
brane comprising the observable universe. We aim to further pursue this idea
in a follow-up publication.

Here we establish how to derive brane quantities
from bulk quantities by adopting Fermi normal coordinates with respect
to the directions parallel to the brane and continuing into the bulk using the
Gauss normal prescription. 
After deriving the equations of motion in the bulk, we project them
parallel and orthogonal to the surface of the brane. 
The brane is assumed to be a distribution of 
$Z_{2}$--symmetric stress-energy about a shell of thickness $2\delta$
in the limit $\delta \to 0.$ Derivatives of quantities discontinuous
across the brane will generate singular distributions on the brane
which relate to the localization of the stress-energy.
This relation is encompassed by the matching conditions across the
brane obtained by the integration of the corresponding equation of
motion in the direction normal to the brane. The matching conditions
provide the boundary conditions on the brane for the 
bulk fields, thus
constraining the parallel projected equations to produce the induced
equations on the brane.
Then we discuss the SSB
mechanism in the context of a braneworld universe embedded in a
five-dimensional anti-de Sitter ($AdS_{5}$) bulk spacetime. 
Spontaneous symmetry breaking is treated by assuming that the
bulk scalar fields acquire a non-vanishing expectation value, which 
induces on the brane 
the breaking of the existing gauge symmetry when the scalar field is
minimally coupled to a gauge field. 

Notice that the formalism employed here has been developed earlier 
\cite{bucher05} and has moreover been used to study a vector field coupled 
non-minimally to gravity in order to examine the implications on the
brane of the spontaneous breaking of Lorentz symmetry in the bulk
\cite{bc06}. In what follows, $d$ stands for the total number of space
dimensions.  

This paper is organized as follows.
In Section \ref{sec:scalar} we work out the case of a bulk scalar
field coupled non-minimally to gravity.
In Section \ref{sec:charged:scalar} we endow the scalar field with a
$U(1)$ charge, introduce an additional gauge field and 
implement the SSB mechanism. This requirement is of particular
relevance since, once the gauge fields  
acquire a mass, they become short range fields and must thus become
confined to the brane. For each case we derive the effective equations of
motion for the bulk fields as induced on the brane.



\section{Scalar Field}
\label{sec:scalar}

In this section we consider a bulk real scalar field $\phi,$ which
contributes to the purely gravitational Lagrangian density for 
$AdS_5$ spacetime
with canonical kinetic and potential terms and with a $\phi$--graviton
interaction term as follows 
\ba
{\cal L}
={1\over {\kappa_{(5)}^2}}R -2\Lambda
+\xi\phi^2R
-{1\over 2}g^{\mu\nu}(\nabla_{\mu}\phi)(\nabla_{\nu}\phi) -V(\phi^2). 
\label{eqn:lagrangian:Phi}
\ea
Here, $\kappa_{(5)}^2=8\pi G_{N(5)}=1/M_{Pl}^3$ is the five-dimensional
gravitational coupling constant  
and $\xi$ is a dimensionless coupling constant 
which measures the non-minimal interaction.
In the cosmological constant term
$\Lambda =\Lambda_{(5)} +\Lambda_{(4)}$ we have included both the bulk
vacuum value $\Lambda_{(5)}$ and that of the brane $\Lambda_{(4)},$
described by a brane tension $\sigma$ localized at the position of the
brane, $\Lambda_{(4)}=\sigma\delta(N).$ 

A distinct feature of our model in comparison with more standard
braneworld approaches 
is the non-minimal coupling of the scalar field to gravity via the
Ricci scalar. As mentioned in the introduction, this type of coupling term
appears in alternative gravity models
and offers an interesting possibility of relating Newton's
constant, the cosmological constant and the mechanism of SSB 
(see e.g. \cite{B87,MCB90} and references therein). 
Given the relevance of these issues, examining their relationship in
the context of the  
braneworlds is quite logical since in these models the Newton's constant, the
five-dimensional cosmological   
constant and the brane tension are entangled. 
In this framework, the mechanism of SSB allows to relate the
mentioned quantities with the  
process of generation of mass.

\subsection{The induced dynamics}

First we derive the equations of motion for both the scalar field and
the graviton as measured by an observer localized on and confined to
the brane. This follows closely the procedure developed in
Ref.~\cite{bc06} for a vector field and in particular uses the results
presented in the appendix therein.

By varying the action with respect to the metric, we obtain the
Einstein equation in the bulk  
\ba
\left( {1\over \kappa_{(5)}^2} +\xi\phi^2\right)G_{\mu\nu} 
+\Lambda g_{\mu\nu}
={1\over 2}T^{(\phi)}_{\mu\nu}
+\xi\Sigma^{(\phi)}_{\mu\nu} ~,
\ea
where
\ba
T^{(\phi)}_{\mu\nu}
=(\nabla_{\mu}\phi)(\nabla_{\nu}\phi)
+g_{\mu\nu}\left[ 
-{1\over 2}g^{\alpha\beta}(\nabla_{\alpha}\phi)(\nabla_{\beta}\phi) 
-V(\phi^2)\right]
\ea
is the stress-energy tensor associated with $\phi$ and
\ba
\Sigma^{(\phi)}_{\mu\nu}
=\nabla_{\mu}\nabla_{\nu}\phi^2  
-g_{\mu\nu}g^{\alpha\beta}\nabla_{\alpha}\nabla_{\beta}\phi^2
\ea
is the contribution from the interaction term.
We note the
$\phi$--dependence of the five-dimensional gravitational coupling
constant akin to that of the Brans-Dicke formulation. 
For the equation of motion for the $\phi$ field, obtained by
varying the action with respect to $\phi,$ we find that
\ba
g^{\mu\nu}\nabla_{\mu}\nabla_{\nu}\phi
-{\partial V\over \partial \phi}
+2\xi\phi R
=0~. 
\ea

We now proceed to project the equations parallel and orthogonal to
the surface of the brane, finding 
for the stress-energy tensor ${\bf T}$ that
\ba
T_{AB}^{(\phi)}&=& (\nabla_{A}\phi)(\nabla_{B}\phi)
+g_{AB}\left[ -{1\over 2}\left[ (\nabla_{C}\phi)^2 +(\nabla_{N}\phi)^2\right]
-V(\phi^2)\right],\cr
T_{AN}^{(\phi)}&=& (\nabla_{A}\phi)(\nabla_{N}\phi)~, \cr
T_{NN}^{(\phi)}&=& (\nabla_{N}\phi)(\nabla_{N}\phi)
+g_{NN}\left[ -{1\over 2}\left[ (\nabla_{C}\phi)^2 +(\nabla_{N}\phi)^2\right] 
-V(\phi^2)\right],
\ea
and similarly 
for the source tensor ${\bf \Sigma}$ that
\ba
\Sigma^{(\phi)}_{AB}
&=&\left( \nabla_{A}\nabla_{B} +K_{AB}\nabla_{N}\right)\phi^2
-g_{AB}\left( \nabla_{C}^2 +\nabla_{N}^2
+K\nabla_{N}\right)\phi^2~, \cr
\Sigma^{(\phi)}_{AN}
&=&\left( \nabla_{A}\nabla_{N} -K_{AB}\nabla_{B}\right)\phi^2~,\cr
\Sigma^{(\phi)}_{NN}
&=&\nabla_{N}\nabla_{N}\phi^2  
-g_{NN}\left( \nabla_{C}^2 +\nabla_{N}^2
+K\nabla_{N}\right)\phi^2~.
\ea
Equating the $(AB)$ components of the decomposition of
the Einstein tensor and of the source terms from the scalar field $\phi,$  
we find for the Einstein equation parallel projected on to the brane that
\ba
&&\left( {1\over \kappa_{(5)}^2} +\xi \phi^2\right)
 \left[ G_{AB}^{(ind)} +2K_{AC}K_{CB} -K_{AB}K -K_{AB,N} 
-{1\over 2}g_{AB}\left(-K_{CD}K_{DC} -K^2 -2K_{,N}\right)\right]
\cr
&=&{1\over 2}(\nabla_{A}\phi)(\nabla_{B}\phi) 
+{1\over 2}g_{AB}\left[ 
-{1\over 2}\left[ (\nabla_{C}\phi)^2 +(\nabla_{N}\phi)^2\right]
-V(\phi^2) \right] - g_{AB}\Lambda\cr
&+&\xi\left[
\left( \nabla_{A}\nabla_{B} +K_{AB}\nabla_{N}\right)\phi^2 
-g_{AB}\left( \nabla_{C}^2 +\nabla_{N}^2 
+K\nabla_{N}\right)\phi^2\right].
\label{eqn:einsteinAB:Phi}
\ea

To obtain the matching condition for the extrinsic curvature across the
brane, we integrate the $(AB)$ component of the Einstein equation in the 
coordinate normal to the brane. 
From the $Z_{2}$--symmetry it follows that $\phi(-\delta)
=\phi(+\delta)$ and consequently that 
$\nabla_{N}\phi(-\delta) =-\nabla_{N}\phi(+\delta).$
Moreover, continuity of the induced metric across the brane
$g_{AB}(-\delta) =g_{AB}(+\delta)$ implies that 
$K_{AB} (-\delta) =-K_{AB} (+\delta).$
We then obtain for the $(AB)$ matching condition 
\ba
&&\int _{-\delta}^{+\delta}dN
\left( {1\over \kappa_{(5)}^2} +\xi \phi ^2\right)
 \left( -K_{AB,N} +g_{AB}K_{,N}\right)\cr
&=&\int _{-\delta}^{+\delta} dN\biggl[
-g_{AB}\Lambda_{(4)}
+\xi \left(
\left( K_{AB} -g_{AB}K\right)\nabla_{N}\phi ^2
-g_{AB}\nabla_{N}^2\phi ^2 \right)\biggr],
\ea
which yields
\ba
\left( {1\over \kappa_{(5)}^2} +\xi \phi ^2\right)
 \left( -K_{AB} +g_{AB}K\right)
= g_{AB}\left( -\sigma -\xi \nabla_{N}\phi ^2\right) 
\label{eqn:imc:Phi}
\ea
for $\phi ^2$ even about the position of the brane.
These provide boundary conditions for ten of the fifteen degrees of
freedom. Five additional boundary conditions are provided by the
matching conditions from the $(AN)$ and $(NN)$ components of the projected
Einstein equations.
From inspection of the $(AN)$ components we note that 
\ba
G_{AN} 
=K_{AB;B} -K_{;A}
&=&-\nabla_{B}\left( \int dN~G_{AB}\right) \cr
&=&-\nabla_{B}\left( \int dN
 {T_{AB}^{(\phi)}\over {1/\kappa _{(5)} ^2 +\xi\phi ^2}}\right) 
=-\nabla_{B}{\cal T}_{AB}^{(\phi)} =0~, 
\ea
which vanishes due to conservation of the induced stress-energy
${\cal T}_{AB}^{(\phi)}$ on the brane, as read off of Eq.~(\ref{eqn:imc:Phi}). 
This condition constrains four degrees of freedom.
The $(NN)$ component of the Einstein equation
\ba
&&\left( {1\over \kappa_{(5)}^2} +\xi \phi ^2\right)\left[
-R^{(ind)} -K_{CD}K_{CD} +K^2\right] +\Lambda _{(5)}\cr
&=&{1\over 2}\left[
{1\over 2}\left( \nabla_{N}\phi\right)\left( \nabla_{N}\phi\right)
-{1\over 2}\left( \nabla_{C}\phi\right)\left( \nabla_{C}\phi\right)
-V(\phi^2)\right]
-\xi\left( \nabla_{C}^2 +K\nabla_{N}\right)\phi ^2
\label{eqn:einsteinNN:Phi}
\ea
consists of the remaining constraint. 
Substituting the $(AB)$ matching condition, Eq.~(\ref{eqn:imc:Phi}),
back in the $(AB)$ Einstein equation, Eq.~(\ref{eqn:einsteinAB:Phi}), we
find for the Einstein equation induced on the brane 
\ba
&&\left( {1\over \kappa_{(5)}^2} +\xi \phi^2\right)
\left[ G_{AB}^{(ind)} +2K_{AC}K_{CB} -K_{AB}K 
-{1\over 2}g_{AB}\left(-K_{CD}K_{DC} -K^2 
\right)\right] +g_{AB}\Lambda_{(5)}\cr
&=&{1\over 2}(\nabla_{A}\phi)(\nabla_{B}\phi) 
+{1\over 2}g_{AB}\left[ 
-{1\over 2}(\nabla_{C}\phi)^2 -{1\over 2}(\nabla_{N}\phi)^2-V\right]
+\xi \left( 
\nabla_{A}\nabla_{B} -g_{AB}\nabla_{C}^2\right)\phi ^2~.\qquad
\label{eqn:einsteinAB:Phi:ind}
\ea

Similarly, we expand the equation of motion for the scalar field
$\phi$ 
\ba
&&\left[ g^{AB}\left( \nabla_{A}\nabla_{B} +K_{AB}\nabla_{N}\right)
+\nabla_{N}\nabla_{N} \right]\phi
-{\partial V\over \partial \phi}\cr
&+&2\xi\phi
 \left( R^{(ind)} -K_{AB}K_{BA} -K^2 -2K_{,N}\right)
=0 ~.
\label{eqn:Phi}
\ea
To obtain the matching condition for $\phi$ across the brane, 
we integrate in the $N$ coordinate discarding all derivatives other
than along $N$ 
\ba
&&\int _{-\delta}^{+\delta}dN\left[
K\nabla_{N}\phi
+\nabla_{N}\nabla_{N}\phi
-4\xi\phi K_{,N}\right] =0~.
\ea
If we assume $Z_{2}$-symmetry across the brane,
the matching condition for $\phi$ becomes
\ba
\nabla_{N}\phi -4\xi K\phi =0 ~.
\label{eqn:Phi:mc}
\ea
Substituting Eq.~(\ref{eqn:Phi:mc}) back into Eq.~(\ref{eqn:Phi}),
we obtain for the propagation of $\phi$ on the brane that
\ba
g^{AB}\nabla_{A}\nabla_{B}\phi
-{\partial V\over \partial \phi}
+2\xi\phi
 \left[ R^{(ind)} -K_{AB}K_{BA} +\left( 1 +8\xi\right)K^2\right]
=0~.
\label{eqn:Phi:ind}
\ea

Moreover, equating the matching conditions for the extrinsic curvature,
Eq.~(\ref{eqn:imc:Phi}), and for the scalar field $\phi,$
Eq.~(\ref{eqn:Phi:mc}), we can solve for $K_{AB}$ and $\nabla_{N}\phi$ 
to find that
\ba
K_{AB}
&=&-g_{AB}~\sigma
{1/(d -1)\over 
 {1/\kappa _{(5)} ^2 +\xi \phi ^2 [1 +8\xi d/(d-1)]}}\vert_{N=0}~,\\
\nabla_{N}\phi\vert_{N=0}
&=&-4\xi\phi~\sigma
{d/(d -1)\over 
 {1/\kappa _{(5)} ^2 +\xi \phi ^2[ 1 +8\xi d/(d -1)]}}\vert_{N=0}~.
\ea
We substitute Eq.~(\ref{eqn:Phi:mc}) for $\nabla_{N}\phi$ and
Eq.~(\ref{eqn:Phi:ind}) for $\nabla_{C} ^2\phi$ in the $(NN)$ component
of the Einstein 
equation, Eq.~(\ref{eqn:einsteinNN:Phi}), to find for $R^{(ind)}$ that
\ba
&&\left( 
{1\over \kappa_{(5)}^2} 
+\xi\phi^2\left( 1 +4\xi\right)\right)R^{(ind)}
=\left( {1\over 4} +2\xi\right)(\nabla_{C} \phi)^2 
+{1\over 2}V
+2\xi\phi{\partial V\over \partial \phi}\cr
&+&\Lambda_{(5)}
-K_{CD}K_{CD}\left( 
{1\over \kappa_{(5)}^2} 
+\xi\phi^2\left( 1 -4\xi\right)\right)
+K^2\left(
{1\over \kappa_{(5)}^2} 
+\xi\phi^2\left( 1 -32\xi ^2\right)\right)~.
\ea
Substituting now in Eq.~(\ref{eqn:Phi:ind}) we find for the equation
of motion for $\phi$ induced on the brane that
\ba
&&g^{AB}\nabla_{A}\nabla_{B}\phi
-{\partial V\over \partial \phi}
\left( {1\over \kappa_{(5)}^2} +\xi \phi^2\right)
 { 1
   \over {1/\kappa_{(5)}^2 +\xi\phi ^2(1 +4\xi)}}\cr
&+&
\left[
\left( {1\over 4} +2\xi\right)\left( \nabla_{C}\phi\right)^2
+{1\over 2}V
+\Lambda_{(5)}\right]
{2\xi\phi\over {1/\kappa _{(5)} ^2 +\xi \phi ^2(1 +4\xi)}}\cr
&+&
K^2
{2\xi\phi\over {1/\kappa _{(5)} ^2 +\xi \phi ^2(1 +4\xi)}}
\left[
{1\over \kappa_{(5)} ^2}\left( -{2\over d} +2 +8\xi\right)
+\xi \phi ^2\left( -{2\over d} +2 +12\xi\right)\right]
=0~.
\label{eqn:Phi:ind2}
\ea
Similarly, substituting in Eq.~(\ref{eqn:einsteinAB:Phi:ind}) we find
for the Einstein equation induced on the brane that
\ba
G_{AB}^{(ind)}
&=&\left( {1\over \kappa_{(5)}^2} +\xi \phi^2\right)^{-1}
 \left[\left( {1\over 2} +2\xi\right)(\nabla_{A}\phi)(\nabla_{B}\phi)  
 +2\xi \phi\nabla_{A}\nabla_{B}\phi\right]\cr
&-&g_{AB}
\left[ 
\left( {1\over 4} +2\xi\right)(\nabla_{C}\phi)^2 
+{1\over 2}V +2\xi \phi {\partial V\over \partial \phi} +\Lambda _{(5)}\right]
  {1\over {1/\kappa _{(5)} ^2 +\xi \phi ^2(1 +4\xi)}}\cr
&-&g_{AB}K^2
\left[
\left( {1\over \kappa_{(5)}^2} +\xi \phi^2\right)
 { { d^2 -d +4}\over {2d^2}}
+(2\xi \phi)^2\left(
{ { -d^2 +3d +4}\over {2d^2}} -8\xi\right)
\right]\cr
&&\times{1\over {1/\kappa _{(5)}^2 +\xi\phi^2(1 +4\xi)}} ~.
\label{eqn:Einstein-phi}
\ea

Using the equations derived above, we realize the case where the scalar
field acquires a non-vanishing vev
$\left< \phi \right>$ which minimizes the
effective potential. This can induce spontaneous symmetry breaking
when the scalar field is coupled to a gauge field, thus endowing the
latter with a mass, 
as will be discussed in Section \ref{sec:charged:scalar}. 
Here, however, a non-vanishing vev will entail a change in the
effective cosmological constant  
and in the effective mass of the scalar field. 
Moreover, once the scalar field acquires a vev, no
direction on the brane can be selected, which implies that 
$\nabla_{A}\left< \phi \right> =0.$ Consequently, Lorentz symmetry breaking 
cannot take place in the presence of a bulk scalar field only (see
Ref. \cite{bc06} for  
the case of an explicit violation of Lorentz symmetry due to a
non-vanishing vev for a vector field). 

We can read off of the induced Einstein equation the effective cosmological
constant, which would 
comprise all the terms proportional to the induced metric which do
not vanish when all the contributions from the matter fields
vanish. 
However, in the case that the matter
fields acquire a non-vanishing vev, the effective
cosmological constant will contain the contribution of the matter
fields at the corresponding non-vanishing value. It follows that
\ba
&&\Lambda _{eff}\left(
{1\over\kappa _{(5)}^2} +\xi \left<\phi\right>^2(1 +4\xi)\right)
=
{1\over 2}V(\left<\phi\right> ^2) 
+2\xi \left<\phi\right> {\partial V\over \partial \phi}
 \Bigg|_{\phi =\left<\phi\right>} 
+\Lambda _{(5)}\cr
&+&K^2|_{\phi =\left<\phi\right>}
\left[
\left( {1\over \kappa_{(5)}^2} +\xi \left<\phi\right>^2\right)
 { { d^2 -d +4}\over {2d^2}}
+(2\xi \left<\phi\right>)^2\left(
{ { -d^2 +3d +4}\over {2d^2}} -8\xi\right)
\right].
\label{eqn:Lambda_eff}
\ea
Moreover, imposing that the effective cosmological vanishes on the
brane, then
\ba
&&\Lambda_{(5)}
=
-{1\over 2}V(\left<\phi\right> ^2)  
-2\xi \left<\phi\right> {\partial V\over \partial \phi}
 \Bigg|_{\phi =\left<\phi\right>}\cr
&-&K^2|_{\phi =\left<\phi\right>}
\left[
\left( {1\over \kappa_{(5)}^2} +\xi \left<\phi\right>^2\right)
 { { d^2 -d +4}\over {2d^2}}
+(2\xi \left<\phi\right>)^2\left(
{ { -d^2 +3d +4}\over {2d^2}} -8\xi\right)
\right].
\label{eqn:Lambda_5}
\ea
We thus observe that a non-vanishing vev in the bulk 
generates 
in the gravitational sector
a contribution to the cosmological constant on the brane.

\subsection{The effective potential}

Whether a non-vanishing vev for the scalar
field can be observed on the brane 
depends on the form of the effective potential $V_{eff}(\phi ^2)$.
The parameters of the potential will influence the magnitude of its minimum
and consequently the mass of the scalar field $\phi$ measured on the brane,
defined as the value of the second derivative of the
effective potential evaluated at the vev of the
scalar field, $\left< \phi \right>$.
We first determine the effective potential measured on the brane and
then proceed to study the conditions for a non-vanishing vev. 

The evolution of $\phi$ on the brane, as described by
Eq.~(\ref{eqn:Phi:ind2}),  
\ba
g^{AB}\nabla_{A}\nabla_{B}\phi 
-{\partial V_{eff}\over \partial \phi}
+\left( {1\over 4} +2\xi\right)\left( \nabla_{C}\phi\right)^2
 {{2\xi \phi}\over {1/\kappa_{(5)} ^2 +\xi\phi ^2(1 +4\xi)}}
=0
\ea
is determined by a damping term as well as by the effective potential
induced on the brane.
Here,
\ba
-{\partial V_{eff}\over \partial \phi}
&=&{1\over {1/\kappa_{(5)} ^2 +\xi\phi ^2(1 +4\xi)}}
\Biggl\{
-{\partial V\over \partial \phi}
 \left( {1\over \kappa_{(5)} ^2} +\xi\phi ^2\right) \cr
&+&
2\xi\phi \left[ 
{1\over 2}V
+\Lambda_{(5)}\right]\cr
&+&
2\xi\phi K^2\left[
{1\over \kappa_{(5)} ^2}\left( -{2\over d} +2 +8\xi\right)
+\xi \phi ^2\left( -{2\over d} +2 +12\xi\right)\right]
\Biggr\} ~,\qquad
\label{eqn:dVeff}
\ea
where $V(\phi ^2)$ is the bulk potential, which is assumed to 
have a Higgs type form
$V(\phi) =\mu _{(5)} ^2 (\phi^2/2) +\lambda _{(5)} (\phi^4/4)$ with
$\lambda_{(5)} >0$. 
We compute $V_{eff}$ by integrating Eq.~(\ref{eqn:dVeff}) 
to find that 
\ba 
V_{eff}(\phi ^2)
&=&
\phi  ^2\left[
{\mu _{(5)}^2\over 2}{1\over {1 +4\xi}}
-{\mu _{(5)}^2\over 4} {1\over {1 +4\xi}}
+\lambda _{(5)}{1 \over {\xi \kappa _{(5)}^2}}{1\over {2( 1 +4\xi)^2}}
 \left( {1\over 4} +4\xi\right)\right]\cr
&+&\phi ^4{\lambda _{(5)}\over 4}\left[
{1\over {1 +4\xi}} 
-{1\over {4(1 +4\xi)}}\right] 
\cr
&+&
\ln\left[ 1  +\xi\kappa _{(5)}^2\phi ^2(1 +4\xi)\right]\times\cr
&&\times\Biggl[ 
\mu _{(5)}^2{1 \over {\xi \kappa _{(5)}^2}}{1\over {2 (1 +4\xi)^2}}
 \left( {1\over 2} +4\xi\right)
-\lambda _{(5)}{1 \over {(\xi\kappa _{(5)}^2)^2}}
  {1\over {2 (1 +4\xi)^3}}
   \left( {1\over 4} +4\xi\right)\cr
&&
-{1\over {1 +4\xi}}
\Lambda _{(5)}
\Biggr]\cr
&+&{\sigma ^2\over 4}{d^2\over {(d +1)(d -1)}}\times\cr
&&\times\Biggl[
{{[ -2/d +2(1 +4\xi)](8d/(d-1)) -4}\over {1 +8\xi d/(d -1)}}
  {1\over { 1/\kappa _{(5)}^2 +\xi\phi ^2[1 +8\xi d/(d -1)]}}\cr
&&-{\kappa _{(5)}^2\over \xi}{{d -1}\over {d +1}}\left(
  -{2\over d} + 1 +8\xi\right)
   \ln\left[ 
    {{ 1 +\xi\kappa _{(5)}^2\phi ^2(1 +4\xi)}\over  
     {{ 1 +\xi\kappa _{(5)}^2\phi ^2[1 +8\xi d/(d -1)]}}}\right]\Biggr]
\ea
for $\xi \not=-1/4.$ 
Notice that, in the limit when $\xi \to 0,$ one recovers the original bulk 
potential with a term on
the brane tension,
$V(\phi ^2) +3\sigma ^2 \kappa _{(5)} ^2d ^2/[(d +1)(d -1)]$. 
It is natural to expect that there exists a hierarchy of scales
depending on whether
the vev $\left< \phi\right>$ is related to the
Standard Model (SM) scale or the grand unified theory scale. Thus, for 
$\vert \xi \phi ^2\vert \ll 1/\kappa_{(5)}^2,$ 
we can expand the denominator of the first term in $\sigma$ 
about $1/\kappa_{(5)}^2,$ keeping terms up to order six in $\phi,$
$(\xi \phi ^2)^3.$ 
The effective potential can thus be written as
\ba
V_{eff}(\phi ^2)
&=& \phi ^2\Biggl\{ 
{\mu _{(5)}^2\over 2}{1\over {1 +4\xi}}
-{\mu _{(5)}^2\over 4} {1\over {1 +4\xi}}
+\lambda _{(5)} {1 \over {\xi \kappa _{(5)}^2}}{1\over {2( 1 +4\xi)}}
 \left( {1\over 4} +4\xi\right)\cr
&&-{1\over 4}\sigma ^2\kappa _{(5)}^2
  (\xi\kappa _{(5)}^2) 
   {d ^2\over {(d +1)(d -1)}}\left[
\left( -{2\over d} +2\left( 1 +4\xi\right)\right)
 {{8d}\over {d -1}} -4\right]
\Biggr\}\cr
&+&\phi ^4 \Biggl\{
{\lambda _{(5)}\over 4}{1\over {1 +4\xi}} 
-{\lambda _{(5)}\over 4}{1\over {4(1 +4\xi)}}\cr
&&+{1\over 4}\sigma ^2\kappa _{(5)}^2
  {1\over 2}(\xi\kappa _{(5)}^2)^2
   {d ^2\over {(d +1)(d -1)}}\left[
\left( -{2\over d} +2\left( 1 + 4\xi\right)\right)
{{8d}\over {d -1}} -4\right]
\left( 1 +\xi{ {8d}\over {d -1}}\right)
\Biggr\}\cr
&-&O[\phi ^6]{1\over 4}\sigma ^2\kappa _{(5)}^2{1\over 6}
({\xi\kappa _{(5)}^2})^3\cr
&+&
\ln\left[ 1 +\xi\kappa _{(5)}^2\phi ^2(1 +4\xi)\right]\times\cr
&&\times\Biggl[ 
\mu _{(5)} ^2{1 \over {\xi \kappa _{(5)}^2}}{1\over {2 (1 +4\xi)^2}}
 \left({1\over 2} +4\xi\right)
-\lambda _{(5)}{1 \over {(\xi\kappa _{(5)}^2})^2}
  {1\over {2 (1 +4\xi)^3}}
   \left( {1\over 4} +4\xi\right)\cr
&&-{1\over {1 +4\xi}}
\Lambda _{(5)}
\Biggr]\cr
&+&{1\over 4}\sigma ^2\kappa _{(5)}^2{d^2\over {(d +1)(d -1)}}\times\cr
&&\times\Biggl[
{{[ -2/d +2(1 +4\xi)](8d/(d-1)) -4}\over {1 +8\xi d/(d -1)}}\cr
&&-{1\over \xi}{{d -1}\over {d +1}}\left(
  -{2\over d} + 1 +8\xi\right)
   \ln\left[ 
    { { 1 +\xi\kappa _{(5)}^2\phi ^2(1 +4\xi)}\over  
     { { 1 +\xi\kappa _{(5)}^2\phi ^2[1 +8\xi d/(d -1)]}}}\right]\Biggr].
\ea
We thus observe that, up to sub-dominant logarithmic terms, the 
effective potential is of the form
$V_{eff} ={\mu _{eff} ^2}(\phi ^2/2) +\lambda_{eff} (\phi ^4/4)+
O[\phi ^6],$ where
\ba
\mu _{eff} ^2 &\sim& 
\mu _{(5)} ^2 
+\lambda_{(5)} {1 \over {\xi \kappa _{(5)}^2}}
-\sigma ^2 \xi \kappa _{(5)}^4~, 
\label{eqn:mu_eff}\\
\lambda _{eff} &\sim&
\lambda _{(5)} 
+\sigma^2 \xi ^2\kappa _{(5)}^6 ~.
\label{eqn:lambda_eff}
\ea 
If $\mu _{eff}^2 <0$ and $\lambda _{eff}>0$, then one expects a
non-vanishing vev for the  
scalar field. The first condition guarantees that a non-vanishing
minimum exists,  
whereas the second condition guarantees that such minimum is finite. 
Conversely, if $\mu _{eff}^2 >0$ and $\lambda _{eff}>0,$ then symmetry
is always unbroken.
Thus, imposing that $\lambda_{eff} >0,$ it follows that 
$\lambda _{(5)} > - \sigma^2 \xi ^2\kappa _{(5)}^6.$
Consequently, in order not to spoil the $\mu _{eff}^2 <0$ condition
we must have that 
$\mu _{(5)}^2 < -2\sigma ^2 \xi \kappa _{(5)}^4.$ 

We notice that the bulk scalar field $\phi,$ being
a five-dimensional field, has dimensions
$[\phi]=M^{3/2}$. Accordingly, $\mu _{(5)}$ has dimensions of mass and
$\lambda _{(5)}$ dimensions of inverse of mass. 
In order to recover characteristically four-dimensional quantities, we
define the four-dimensional scalar field $\Phi$ as the rescaling of
$\phi$ by an appropriate mass scale $M_{(\phi)}$. 
In the mode expansion of a bulk field, this mass can be identified
with the mode function dependent on the direction $N$ evaluated at the
position of the brane in the bulk.
Thus, for $\phi =M_{\phi}^{{1\over 2}}\Phi,$
the induced equation of motion for $\Phi$ on the brane becomes
\ba
g^{AB}\nabla_{A}\nabla_{B}\Phi 
-{1\over M_{\phi}}{\partial V_{eff}\over \partial \Phi} 
+ \left( {1\over 4} +2\xi\right)\left( \nabla_{C}\Phi\right)^2 
 {{2\xi M_{\phi} \Phi}\over 
  {1/\kappa_{(5)} ^2 +\xi M_{\phi}\Phi ^2(1 +4\xi)}} 
=0 ~.
\ea
Consequently, the parameters of the effective potential will scale as
\ba
{1\over M _{\phi}}V_{eff}(\Phi ^2)
=\mu _{eff} ^2 \Phi ^2 +\lambda _{eff}M _{\phi}\Phi ^4 
+M _{\phi} ^2 O[\Phi ^6]~.
\ea
with equations (\ref{eqn:mu_eff}) and (\ref{eqn:lambda_eff}) becoming
\ba
\mu _{eff} ^2 &\sim& 
\mu ^2 
-2\sigma ^2 \xi \kappa _{(5)}^4~,\\ 
M_{\phi}\lambda _{eff} &\sim&
\lambda 
+M_{\phi}\xi^2 \sigma^2 \kappa _{(5)}^6 ~,
\ea
where $\mu =\mu _{(5)}$ and $\lambda =M_{\phi}\lambda _{(5)}.$
Here, for $\xi >0$ we have two possible mechanisms for the generation of 
a non-vanishing vev: the canonical way, via the potential
associated with the scalar field, and the braneworld way, via the
interaction of the scalar field with the brane tension.
For the latter to be viable in the context of the SM, then 
\ba
\left|{\mu _{eff} ^2\over {M_{\phi}\lambda _{eff} }}\right|
\sim {1\over {\xi M_{\phi}}} {1\over \kappa _{(5)}^2}
\ea
must be of order $TeV ^2$, and $\left< \Phi\right> = 246$ GeV.
However, in order to recover the four-dimensional gravitational
coupling constant in Eq.~(\ref{eqn:Einstein-phi}), we find from the
$\phi$ contribution that 
$
M_{Pl(4)} ^{-2} =\kappa _{(5)}^2M_{\phi}
$
and from the $\sigma$ contribution that
$
M_{Pl(4)} ^{-2} \sim\kappa _{(5)}^4\sigma.
$
It follows that
\ba
M_{Pl} ^3\equiv {1\over \kappa _{(5)}^2}
\sim{\sigma \over M_{\phi}}
\label{eqn:Planck_5:phi}
\ea
with $\sigma \sim M_{Pl(4)} ^2 M_{\phi}^2.$ This implies that
\ba
\left|{\mu _{eff} ^2\over {M_{\phi}\lambda _{eff} }}\right|
\sim {1\over {\xi}}M_{Pl(4)} ^2 \gg TeV ^2,
\label{eqn:Planck_4}
\ea
which renders the brane mediated mechanism of SSB 
unviable for phenomenological reasons. 
This means that the phenomenological hierarchy between the SM 
typical energy scale order $TeV$ and the Planck scale of the 
gravitational effects of the physics on the brane cannot be powered
by the brane mediated mechanism of SSB, since the characteristic scale
of the induced dynamics of the scalar field is the Planck scale.
It is easy to see that $\left< \Phi\right> \sim M_{Pl(4)}.$ Thus, the
scalar field becomes a short range  
field about the brane and therefore strongly localized therein.

Moreover, from the expression for the five-dimensional cosmological in the case
of a vanishing effective cosmological constant, Eq.~(\ref{eqn:Lambda_5}), 
we find that $\Lambda _{(5)} \sim -M_{\phi}V ( \left< \Phi \right> ^2) 
-\sigma ^2 \kappa _{(5)}^2 $, 
and consequently that 
\ba
M_{Pl} ^3\equiv {1\over \kappa _{(5)}^2} 
\sim {\sigma ^2\over{-\Lambda _{(5)} -M_{\phi}V ( \left< \Phi \right> ^2)}} ~.
\label{eqn:Planck_5:Lambda_5}
\ea
It then follows that $-\Lambda _{(5)} -M_{\phi}V ( \left< \Phi \right> ^2)
\sim M_{Pl(4)} ^2M_{\phi} ^3$.

The vev of $\phi$
is localized at the minimum of the effective
potential, $(\partial V_{eff}/\partial \phi)=0.$
Only real and positive solutions for $\left< \phi\right>$ are physically
relevant. 
It is possible to show that, besides $\left< \Phi\right> =0$, 
of the four real solutions only one is positive for
all values of the equation parameters 
is hence physically acceptable.

\section{Charged Scalar Field}
\label{sec:charged:scalar}

In this section, in order to study spontaneous symmetry breaking, 
we introduce a minimal coupling of a
scalar field $\psi$ to a bulk $U(1)$ gauge potential ${\bf B}$
through the covariant derivative 
$\tilde \nabla_{\mu} =\nabla_{\mu} +ieB_{\mu}.$
However, only the scalar field is coupled non-minimally to the graviton. 
For the invariance of the action under local gauge transformations,
the scalar field $\psi$ must be complex.
The coupling constant $e$ is a measure
of the charge of the scalar field associated with the field
generated by the gauge potential. In order to avoid that flux of the
gauge field  
leaks to the bulk spacetime, the gauge symmetry must be broken and the
field thus confined to the brane. 
However, the dynamics of the symmetry breaking cannot be fully
understood before  
properly treating the field in the bulk and studying 
how it appears from the point of view of the brane. 

The Lagrangian density consists of the terms introduced in the
previous section, Eq.~(\ref{eqn:lagrangian:Phi}), and additionally of
the kinetic term associated with the gauge potential, given as follows
\ba
{\cal L} 
={1\over {\kappa_{(5)}^2}}R -2\Lambda
+\xi\psi^2R
- g^{\mu\nu}(\tilde \nabla_{\mu}\psi)(\tilde \nabla_{\nu}\psi^{\ast}) 
-V(\psi^2)
-{1\over 4}B_{\mu\nu}B^{\mu\nu}~.
\ea
Here, $\psi^2 = \psi \psi^{\ast}$ and
$B_{\mu\nu} 
=\partial_{\mu}B_{\nu} -\partial_{\nu}B_{\mu}$ is the field strength
tensor associated with ${\bf B}.$ 

\subsection{The induced dynamics}
First we work out how this bulk theory is observed on the brane.
By varying the action with respect to the metric, we 
obtain the Einstein equation in the bulk
\ba
\left( {1\over \kappa_{(5)}^2} +\xi\psi^2\right)G_{\mu\nu} 
+\Lambda g_{\mu\nu}
={1\over 2}T_{\mu\nu}^{(\psi)}
+{1\over 2} T_{\mu\nu}^{(B)}
+\xi\Sigma^{(\psi)}_{\mu\nu}~, 
\label{eqn:Einstein:Psi}
\ea
where
\ba
T_{\mu\nu}^{(\psi)}
&=&2(\tilde \nabla_{\mu}\psi)(\tilde \nabla_{\nu}\psi ^{\ast})
+g_{\mu\nu}\left[ 
-g^{\alpha\beta}
 (\tilde \nabla_{\alpha}\psi)(\tilde \nabla_{\beta}\psi ^{\ast}) 
-V(\psi^2)\right], \\
T_{\mu\nu}^{(B)} 
&=&B_{\mu\rho}B_{\nu}{}^{\rho} 
-{1\over 4}g_{\mu\nu}B_{\rho\sigma}B^{\rho\sigma} 
\ea
are the stress-energy tensors associated with the fields 
$\psi$ and ${\bf B}$ respectively, 
and 
\ba
\Sigma_{\mu\nu}^{(\psi)}
&=&
\psi\nabla_{\mu}\nabla_{\nu}\psi ^{\ast} 
+\psi ^{\ast}\nabla_{\mu}\nabla_{\nu}\psi
+(\nabla_{\mu}\psi)(\nabla_{\nu}\psi ^{\ast})
+(\nabla_{\mu}\psi ^{\ast})(\nabla_{\nu}\psi)\cr
&-&g_{\mu\nu}g^{\alpha\beta}\left[
\psi\nabla_{\alpha}\nabla_{\beta}\psi ^{\ast}
+\psi ^{\ast}\nabla_{\alpha}\nabla_{\beta}\psi
+2(\nabla_{\alpha}\psi)(\nabla_{\beta}\psi ^{\ast})
\right]
\ea
is the contribution from the $\psi$--graviton interaction term.

We now proceed to derive the induced Einstein equation.
First we project the source terms in Eq.~(\ref{eqn:Einstein:Psi})
along the parallel and orthogonal directions to the surface of the
brane. The components are as follows  
for $T_{\mu\nu}^{(\psi)}$ 
\ba
T_{AB}^{(\psi)}
&=&2(\nabla_{A} +ieB_{A})\psi (\nabla_{B} -ieB_{B})\psi ^{\ast}\cr
&-&g_{AB}\left[
(\nabla_{C} +ieB_{C})\psi (\nabla_{C} -ieB_{C})\psi ^{\ast}
+(\nabla_{N} +ieB_{N})\psi (\nabla_{N} -ieB_{N})\psi ^{\ast}
+V\right]~,\cr
T_{AN}^{(\psi)}
&=&2(\nabla_{A} +ieB_{A})\psi (\nabla_{N} -ieB_{N})\psi ^{\ast}~,\cr
T_{NN}^{(\psi)}
&=&
-(\nabla_{C} +ieB_{C})\psi (\nabla_{C} -ieB_{C})\psi ^{\ast}
+(\nabla_{N} +ieB_{N})\psi (\nabla_{N} -ieB_{N})\psi ^{\ast}
-V~,
\ea
for $T_{\mu\nu}^{(B)}$
\ba
T_{AB}^{(B)}
&=&(\nabla_{A}B_{C} -\nabla_{C}B_{A})(\nabla_{B}B_{C} -\nabla_{C}B_{B})
+(\nabla_{A}B_{N} -\nabla_{N}B_{A})(\nabla_{B}B_{N} -\nabla_{N}B_{B})\cr
&-&{1\over 4}g_{AB}\left[
(\nabla_{C}B_{D} -\nabla_{D}B_{C})^2 
+2(\nabla_{N}B_{D} -\nabla_{D}B_{N})^2\right] ~,\cr
T_{AN}^{(B)}
&=&(\nabla_{A}B_{C} -\nabla_{C}B_{A})(\nabla_{N}B_{C} -\nabla_{C}B_{N})~,\cr
T_{NN}^{(B)}
&=&
-{1\over 4}\left[
(\nabla_{C}B_{D} -\nabla_{D}B_{C})^2 
-2(\nabla_{N}B_{D} -\nabla_{D}B_{N})^2\right]~,
\ea
and for $\Sigma_{\mu\nu}^{(\psi)}$
\ba
\Sigma_{AB}^{(\psi)}
&=&
\psi(\nabla_{A}\nabla_{B} +K_{AB}\nabla_{N})\psi ^{\ast}
+\psi ^{\ast}(\nabla_{A}\nabla_{B} +K_{AB}\nabla_{N})\psi
+(\nabla_{A}\psi)(\nabla_{B}\psi ^{\ast})
+(\nabla_{A}\psi ^{\ast})(\nabla_{B}\psi) \cr
&-&g_{AB}\Bigl[
\psi \left( \nabla_{C}^2 +K\nabla_{N} +\nabla_{N}^2\right)\psi ^{\ast} 
+\psi ^{\ast} \left( \nabla_{C}^2 +K\nabla_{N} +\nabla_{N}^2\right)\psi \cr
&&+2(\nabla_{C}\psi)(\nabla_{C}\psi ^{\ast}) 
+2(\nabla_{N}\psi)(\nabla_{N}\psi ^{\ast})\Bigr]~,\cr
\Sigma_{AN}^{(\psi)}
&=& 
\psi(\nabla_{A}\nabla_{N} -K_{AB}\nabla_{B})\psi ^{\ast}
+\psi ^{\ast}(\nabla_{A}\nabla_{N} -K_{AB}\nabla_{B})\psi
+(\nabla_{A}\psi)(\nabla_{N}\psi ^{\ast})
+(\nabla_{A}\psi ^{\ast})(\nabla_{N}\psi)~, \cr
\Sigma_{NN}^{(\psi)}
&=& -\left[
\psi \left( \nabla_{C}^2 +K\nabla_{N}\right)\psi ^{\ast} 
+\psi ^{\ast} \left( \nabla_{C}^2 +K\nabla_{N}\right)\psi 
+2(\nabla_{C}\psi)(\nabla_{C}\psi ^{\ast})\right]~.
\ea
We equate the $(AB)$ components of the decomposition of the Einstein
tensor and of the source terms from both the scalar field and the
gauge potential to obtain the  $(AB)$ component of the Einstein
equation
\ba
&&\left( {1\over \kappa_{(5)}^2} +\xi \psi^2\right)
 \left[ G_{AB}^{(ind)} +2K_{AC}K_{CB} -K_{AB}K -K_{AB,N} 
-{1\over 2}g_{AB}\left(-K_{CD}K_{DC} -K^2 -2K_{,N}\right)\right]
\cr
&+&g_{AB}\Lambda
={1\over 2}T_{AB}^{(\psi)}+{1\over 2}T_{AB}^{(B)}+\xi \Sigma_{AB} ^{(\psi)}~.
\ea
Next we proceed to integrate the $(AB)$ component of the Einstein
equation in the coordinate normal to the brane to obtain the matching
conditions for the extrinsic curvature across the brane.
We find that 
\ba
\left[ 
\left( {1\over \kappa_{(5)}^2} +\xi\psi^2\right)
 \left( -K_{AB} +g_{AB}K\right)
\right]^{+\delta}_{-\delta}
&=&\int ^{+\delta}_{-\delta}dN \left[-g_{AB}\Lambda_{(4)}\right]
-g_{AB}\xi \left[
\psi\nabla_{N}\psi ^{\ast}
+\psi ^{\ast}\nabla_{N}\psi 
\right]^{+\delta}_{-\delta}~,\qquad
\ea
which yields
\ba 
\left( {1\over \kappa_{(5)}^2} +\xi\psi^2\right)
 \left( -K_{AB} +g_{AB}K\right)
=-g_{AB}\left[ \sigma
+\xi \left(
 \psi\nabla_{N}\psi ^{\ast} 
 +\psi ^{\ast}\nabla_{N}\psi \right)
\right]~.
\label{eqn:imc:Psi}
\ea
These provide ten boundary conditions for the extrinsic curvature on
the brane.
As noticed in the previous section, equating the $(AN)$ components of
the Einstein equation we find four constraint conditions on the
extrinsic curvature, $K_{AB;B} -K_{;A}=0,$
which follow from the conservation of the
stress-energy tensor induced on the brane.
The $(NN)$ equation 
\ba
\left( {1\over \kappa_{(5)}^2} +\xi\psi^2\right)
 \left[ -R^{(ind)} -K_{CD}K_{CD} +K^2\right] 
+\Lambda _{(5)}
={1\over 2}T_{NN}^{(\psi)}+{1\over 2}T_{NN}^{(B)}+\xi \Sigma_{NN} ^{(\psi)}
\label{eqn:GNN}
\ea
provides the remaining condition relating the intrinsic curvature of
the induced metric with the stress-energy sources. 

To derive the equation of motion for the scalar field we 
vary the action with respect to $\psi$ and $\psi ^{\ast},$ finding 
respectively for $\psi ^{\ast}$ the equation 
\ba
g^{\mu\nu}\left[
\nabla_{\mu}\nabla_{\nu} -2ieB_{\mu}\nabla_{\nu} +(ieB_{\mu})(ieB_{\nu})
-ie(\nabla_{\mu}B_{\nu})\right]\psi ^{\ast}
-{\partial V\over \partial \psi} +\xi\psi ^{\ast}R =0~, 
\label{eqn:Psi^ast}
\ea
and for $\psi$ the corresponding complex conjugate equation
\ba
g^{\mu\nu}\left[
\nabla_{\mu}\nabla_{\nu} +2ieB_{\mu}\nabla_{\nu} +(ieB_{\mu})(ieB_{\nu})
+ie(\nabla_{\mu}B_{\nu})\right]\psi 
-{\partial V\over \partial \psi ^{\ast}} +\xi\psi R =0 ~.
\label{eqn:Psi}
\ea
We rewrite these equations by expanding the tensor quantities as
follows
\ba
&&\biggl[
g^{AB}\left( \nabla_{A}\nabla_{B} +K_{AB}\nabla_{N}\right)
+\nabla_{N}\nabla_{N}
-2ie\left( g^{AB}B_{A}\nabla_{B} +B_{N}\nabla_{N}\right)\cr
&+&g^{AB}(ieB_{A})(ieB_{B}) +(ieB_{N})(ieB_{N})
-ieg^{AB}\left( \nabla_{A}B_{B} +K_{AB}B_{N}\right) 
-ie\nabla_{N}B_{N}\biggr]\psi ^{\ast}\cr
&-&{\partial V\over \partial \psi} 
+\xi\psi ^{\ast}\left( R^{(ind)} -K_{AB}K_{BA} -K^2 -2K_{,N}\right) =0 ~,
\ea
and equivalently for the corresponding complex conjugate.
We now integrate in the $N$ direction to find the matching condition for
$\psi ^{\ast}$ 
\ba
\int ^{+\delta} _{-\delta}dN\left[
\left( 
K\nabla_{N} 
+\nabla_{N}\nabla_{N}
-ie B_{N}\nabla_{N} -ie \nabla_{N}B_{N}\right)\psi ^{\ast}
-2\xi \psi ^{\ast}K_{,N}\right] =0,
\ea
which yields
\ba
\left[ 
-2\xi K\psi ^{\ast}
+(\nabla_{N}
-ieB_{N})\psi ^{\ast}\right]^{+\delta} _{-\delta} =0 ~.
\label{eqn:Psi^ast:mc}
\ea
Substituting back into Eq.~(\ref{eqn:Psi^ast})
and eliminating the $\nabla_{N}\psi ^{\ast}$ term, we obtain
\ba
&&g^{AB}\left[
\nabla_{A}\nabla_{B} 
-2ie B_{A}\nabla_{B} 
+(ieB_{A})(ieB_{B})
-ie\nabla_{A}B_{B} \right]\psi ^{\ast}\cr
&-&{\partial V\over \partial \psi} 
+\xi\psi ^{\ast}\left( R^{(ind)} -K_{AB}K_{BA} +(1 +4\xi)K^2\right) =0 ~,
\ea
where the extrinsic curvature results from the matching condition in
Eq.~(\ref{eqn:imc:Psi}) 
\ba
K_{AB}
&=&-g_{AB}~\sigma
{1/(d -1)\over 
 {1/\kappa _{(5)} ^2 +\xi \psi ^2 (1 +4\xi d/(d-1))}}
\label{eqn:K}
\ea
and the intrinsic curvature results from the $(NN)$ component of the Einstein
equation in Eq.~(\ref{eqn:GNN})
\ba
&&\left( 
{1\over \kappa_{(5)}^2} +\xi\psi^2( 1 +2\xi) \right)R^{(ind)}
=\left( {1\over 2} +2\xi\right)
  (\nabla_{C} +ieB_{C})\psi (\nabla_{C} -ieB_{C})\psi ^{\ast} \cr
&+&{1\over 2}V
+\xi\left( 
\psi{\partial V\over \partial \psi}
+\psi ^{\ast}{\partial V\over \partial \psi ^{\ast}}\right)
+{1\over 8}\left(\nabla_{C}B_{D} -\nabla_{D}B_{C}\right)\cr
&+&\Lambda_{(5)}
-K_{CD}K_{CD}\left( 
{1\over \kappa_{(5)}^2} 
+\xi\psi^2( 1 -2\xi)\right)
+K^2 \left(
{1\over \kappa_{(5)}^2} +\xi\psi^2( 1 -8\xi ^2) \right) ~.
\ea
The matching condition and the induced equation of motion for $\psi$
are the complex conjugate of the corresponding equations for 
$\psi ^{\ast}.$ 

For the equation of motion for the vector field ${\bf B},$ obtained by
varying the action with respect to $B_{\mu},$ we find that
\ba
\nabla^{\nu}
 \left( \nabla_{\nu}B_{\mu} -\nabla_{\mu}B_{\nu}\right)
+ieg_{\mu\nu}\left( 
\psi ^{\ast}\tilde \nabla ^{\nu}\psi 
-\psi \tilde \nabla ^{\nu}\psi ^{\ast}\right)
=0 ~.
\ea
We project the equation parallel and orthogonal to
the surface of the brane 
to find for the parallel component 
that
\ba
&&\nabla_{C}( \nabla_{C}B_{A} -\nabla_{A}B_{C})
+\nabla_{N}( \nabla_{N}B_{A} -\nabla_{A}B_{N})\cr
&+&2K_{AC}( \nabla_{C}B_{N} -\nabla_{N}B_{C})
+K( \nabla_{N}B_{A} -\nabla_{A}B_{N})\cr
&+&ie\left( 
\psi ^{\ast}\nabla _{A}\psi 
-\psi \nabla _{A}\psi ^{\ast}\right) +(ie)^2B_{A}\psi ^2
=0~,
\label{eqn:B_A}
\ea
and for the orthogonal component 
that
\ba
\nabla_{C}( \nabla_{C}B_{N} -\nabla_{N}B_{C})
+ie\left( 
\psi ^{\ast}\nabla _{N}\psi 
-\psi \nabla _{N}\psi ^{\ast}\right) +(ie)^2B_{N}\psi ^2
=0 ~.
\label{eqn:B_N}
\ea
Integrating Eqs.~(\ref{eqn:B_A}) and (\ref{eqn:B_N}) in the $N$
direction, we
find the matching conditions respectively for $B_{A}$ 
\ba
\left[ \nabla_{N}B_{A} -\nabla_{A}B_{N}\right]^{+\delta} _{-\delta}=0~,
\label{eqn:B_A:mc}
\ea
and for $B_{N}$
\ba
\left[ \nabla_{C}B_{C} -KB_{N}\right]^{+\delta} _{-\delta}=0 ~.
\label{eqn:B_N:mc}
\ea
We then substitute Eq.~(\ref{eqn:B_A:mc}) in Eq.~(\ref{eqn:B_A}),
obtaining for the induced $B_{A}$ equation
\ba
\nabla_{C}( \nabla_{C}B_{A} -\nabla_{A}B_{C})
+ie\left( 
\psi ^{\ast}\nabla _{A}\psi 
-\psi \nabla _{A}\psi ^{\ast}\right) +(ie)^2B_{A}\psi ^2
=0 ~.
\label{eqn:B_A_eff}
\ea
We notice that the component $B_A$ will
acquire on the brane a mass proportional to the scalar field vev. 
When $\psi$ acquires a non-vanishing vev, gauge invariance is
spontaneously broken on the brane for $B_{A},$ which thus 
becomes a short range field about the brane. 

To derive the induced $B_{N}$ equation we similarly substitute
Eq.~(\ref{eqn:B_N:mc}) in Eq.~(\ref{eqn:B_N}) and moreover use the
matching conditions for the scalar field, Eq.~(\ref{eqn:Psi^ast:mc}) and
its complex conjugate, obtaining
\ba
\nabla_{C}\nabla_{C}B_{N} 
-KB_{N,N} -K_{CD}K_{DC}B_{N} 
=0 ~.
\label{eqn:B_N_eff}
\ea
We notice that on the brane 
$B_{N}$ does not interact with the scalar field but only with the
brane tension, which
is manifest in the vanishing coupling constant to $\psi$ 
induced on the brane. Consequently, in the case where $\psi$
acquires a non-vanishing vev, $B_{N}$ does not
undergo spontaneous symmetry breaking, thus remaining massless with
respect to $\psi.$ 
(The breaking of gauge symmetry for $B_{N}$ would require a
non-minimal coupling  of ${\bf B}$
to gravity, as discussed in Ref. \cite{bc06}.) However, the $B_{N}$
component acquires a  
mass with respect to the brane tension $\sigma,$ 
which implies that $B_N$ is strictly localized on the brane.

Finally, we substitute the matching conditions for the extrinsic
curvature, Eq.~(\ref{eqn:imc:Psi}), 
in the $(AB)$ component of the Einstein equation and moreover use the
matching conditions for the scalar field, Eq.~(\ref{eqn:Psi^ast:mc}),
as well as for the vector field, 
Eqs.~(\ref{eqn:B_A:mc}) and (\ref{eqn:B_N:mc}), to
find for the Einstein equation induced on the brane that
\ba
&&\left( {1\over \kappa_{(5)}^2} +\xi\psi^2\right)
G_{AB} ^{(ind)}= \cr
&=&
\xi\Bigl[
\psi \nabla_{A}\nabla_{B}\psi ^{\ast}
+\psi ^{\ast}\nabla_{A}\nabla_{B}\psi 
+( \nabla_{A}\psi)( \nabla_{B}\psi ^{\ast})
+( \nabla_{A}\psi ^{\ast})( \nabla_{B}\psi) \cr
&&-g_{AB}g^{CD}\left(
\psi \nabla_{C}\nabla_{D}\psi ^{\ast}
+\psi ^{\ast}\nabla_{C}\nabla_{D}\psi 
+2( \nabla_{C}\psi)( \nabla_{D}\psi ^{\ast})
\right)\Bigr]\cr
&+&\biggl[
(\nabla_{A} +ieB_{A})\psi (\nabla_{B} -ieB_{B})\psi ^{\ast}\cr
&&-{1\over 2}g_{AB}\left(
 g^{CD}(\nabla_{C} +ieB_{C})\psi (\nabla_{D} -ieB_{D})\psi ^{\ast}
+{1\over 2V}+\Lambda_{(5)}\right)\biggr]\cr
&+&{1\over 2} \biggl[
(\nabla_{A}B_{C} -\nabla_{C}B_{A})(\nabla_{B}B_{C} -\nabla_{C}B_{B})\cr
&&-g_{AB}{1\over 4}
 (\nabla_{C}B_{D} -\nabla_{D}B_{C})(\nabla_{C}B_{D} -\nabla_{D}B_{C})
\biggr]\cr
&-&g_{AB}K^2 \left( {1\over \kappa_{(5)}^2} +\xi\psi^2\right)
\left\{ {{d ^2 -d +4}\over {2d ^2}} +2\xi ^2\psi ^2\right\},
\ea
which, upon substitution of the induced equations of motion for the scalar
field, further reduces to 
\ba
&&
G_{AB} ^{(ind)}=\cr
&=&\left( {1\over \kappa_{(5)}^2} +\xi\psi^2\right)^{-1}\Biggl[
 \left( \nabla_{A} +ieB_{A}\right)\psi
  \left( \nabla_{B} -ieB_{B}\right)\psi ^{\ast}
+{1\over 2}
 \left( \nabla_{A}B_{C} -\nabla_{C}B_{A}\right)
  \left( \nabla_{B}B_{C} -\nabla_{C}B_{B}\right)\cr
&&+\xi\left[
\psi \nabla_{A}\nabla_{B}\psi ^{\ast}
+\psi ^{\ast} \nabla_{A}\nabla_{B}\psi
+\left( \nabla_{A}\psi\right)\left( \nabla_{B}\psi ^{\ast}\right)
 +\left( \nabla_{A}\psi ^{\ast}\right)\left( \nabla_{B}\psi \right)
\right]\Biggr]\cr
&-&g_{AB}\Biggl[
\left( {1\over 2} +2\xi\right)
 \left( \nabla_{C} +ieB_{C}\right)\psi
  \left( \nabla_{C} -ieB_{C}\right)\psi ^{\ast}
+{1\over 8}\left( \nabla_{D}B_{C} -\nabla_{C}B_{D}\right)^2 \cr
&&+{1\over 2}V +\Lambda_{(5)} 
+\xi\left( \psi {\partial V\over \partial \psi} 
+\psi ^{\ast} {\partial V\over \partial \psi ^{\ast}}\right)
\Biggr]{1\over {1/\kappa_{(5)}^2 +\xi\psi ^2(1 +2\xi)}}\cr
&-&
g_{AB}K^2
\left[
\left( {1\over \kappa_{(5)}^2} +\xi\psi^2\right)
 {{d ^2 -d +4}\over {2d ^2}}
+2\xi^2\psi ^2\left( {{-d ^2 +3d +4}\over {2d ^2}} -4\xi\right)\right]\cr
&&\times {1\over {1/\kappa_{(5)}^2 +\xi\psi ^2(1 +2\xi)}} ~.
\label{eqn:Einstein-psi}
\ea

We now proceed to realize the case where the scalar field $\psi$ 
acquires a non-vanishing vev
$\left< \psi\right>,$ with $\psi ^{\ast}$ acquiring the corresponding
conjugate value $\left<\psi\right> ^{\ast}.$
In order to avoid the breaking of
Lorentz symmetry, we must have that 
$\nabla_{A}\left< \psi\right>=\nabla_{A}\left< \psi\right>^{\ast} =0$
and $\left< {\bf B}\right> =0.$ 
Moreover, $\left< {\bf B}\right>$
must not be allowed to vary in space but must instead be forced 
to take on the same value everywhere in the bulk, 
which implies both 
$\nabla _{A} \left< {\bf B}\right> =0$ and  
$\nabla _{N} \left< {\bf B}\right> =0.$

We can read off of the induced Einstein equation the effective
cosmological constant on the brane, defined as the term proportional
to the induced metric
\ba
&&\Lambda _{eff}\left(
{1\over\kappa _{(5)}^2} 
+\xi\left<\psi\right>^2(1 +2\xi)\right)
={1\over 2}V(\left<\psi\right>^2) +\Lambda_{(5)} 
+\xi\left( 
\left<\psi\right> 
 {\partial V\over \partial \psi}\Bigg|_{\psi =\left<\psi\right>} 
+\left<\psi\right> ^{\ast} 
 {\partial V\over \partial \psi ^{\ast}}\Bigg|_{\psi =\left<\psi\right>}\right)
\cr
&+&K^2\Big|_{\psi =\left<\psi\right>}
\left[
\left( {1\over \kappa_{(5)}^2} +\xi\left<\psi\right>^2\right)
 {{d ^2 -d +4}\over {2d ^2}}
+2\xi^2\left<\psi\right>^2\left( 
{{-d ^2 +3d +4}\over {2d ^2}} -4\xi\right)\right].
\label{eqn:Lambda_eff_psi}
\ea

\subsection{The effective potential}

We now proceed to determine the effective potential for the field
$\psi$ and study the conditions for SSB,
following the same reasoning as that for the real scalar field $\phi$.

The evolution of $\psi ^{\ast}$ on the brane is described by
\ba
&&g^{AB}\nabla_{A}\nabla_{B} \psi ^{\ast} 
-{\partial V_{eff}\over \partial \psi}\cr
&+&\Biggl[
-(ieB_{C})\nabla_{C}\psi ^{\ast}\left(
2{1\over \kappa _{(5)} ^2} 
+\xi\psi ^2\left({3\over 2} +2\xi\right)\right)\cr
&&+\xi\psi ^{\ast}\left[
\left( {1\over 2} +2\xi\right)\left\{
\left( \nabla_{C}\psi\right)\left( \nabla_{C}\psi ^{\ast}\right)
-\left( ieB_{C}\right)\psi ^{\ast}\left( \nabla_{C}\psi\right)\right\}
\right]
\Biggr]{1\over {1/\kappa _{(5)} ^2 +\xi \psi ^2( 1 +2\xi)}}
=0~,\qquad
\ea
where 
\ba
&-&{\partial V_{eff}\over \partial \psi}
=-ie\left( \nabla_{C}B_{C}\right)\psi ^{\ast}\cr
&+&\Biggl\{
-{\partial V\over \partial \psi}\left(
{1\over \kappa _{(5)} ^2} +\xi\psi ^2( 1 +\xi)\right)
+\xi ^2\psi ^{\ast}\psi ^{\ast}{\partial V\over \partial \psi ^{\ast}}
\cr
& 
+&(ieB_{C})^2\psi ^{\ast}\left(
{1\over \kappa _{(5)} ^2} +{1\over 2}\xi\psi ^2\right)\cr
&+&\xi\psi ^{\ast}\left[
{1\over 8}\left(\nabla_{C}B_{D} -\nabla_{D}B_{C}\right)^2
+{1\over 2}V +\Lambda_{(5)}\right]\cr
&+&\xi\psi ^{\ast}K ^2\left[
{1\over \kappa _{(5)} ^2}\left(
-{2\over d} +2 +4\xi\right)
+\xi \psi ^2\left(
-{2\over d} +2 +6\xi\right)\right]
\Biggr\}{1\over {1/\kappa _{(5)} ^2 +\xi \psi ^2( 1 +2\xi)}}.
\ea
Assuming as before that 
$V(\psi\psi ^{\ast} ) =\mu _{(5)} ^2 \psi\psi ^{\ast} 
+ \lambda _{(5)} (\psi\psi ^{\ast})^2,$ 
we integrate in $\psi$ to
find for the effective potential
\ba
V_{eff}(\psi ^2)
&=&\psi ^2\Biggl[
{\mu _{(5)} ^2\over 2}{1\over { 1 +2\xi}}
-{\mu _{(5)} ^2\over 4}{1\over { 1 +2\xi}} 
+\lambda _{(5)}{1\over {\xi\kappa _{(5)} ^2}}{1\over {2( 1 +2\xi)^2}}
  \left( {1\over 4} +2\xi\right)\cr
&&-(ieB_{C})^2 {1\over {2( 1 +2\xi)}}
+ie(\nabla_{C}B_{C})\Biggr]\cr
&+&\psi ^4{\lambda _{(5)}\over 4}\left[
{1\over {1 +2\xi}} -{1\over {4( 1 +2\xi)}}\right]
\cr
&+&\ln\left[ 1 +\psi ^2\kappa _{(5)} ^2\xi (1 +2\xi)\right]
\times \cr
&&\times
\Biggl[ 
\mu _{(5)} ^2 {1\over {\xi \kappa _{(5)}^2}} {1\over {2(1 +2\xi) ^2}}
 \left( {1\over 2} +2\xi\right)
-\lambda _{(5)} {1\over {( \xi \kappa _{(5)}^2})^2}
 {1\over {2(1 +2\xi)^3}}\left( {1\over 4} +2\xi\right)\cr
&&-{1\over {\xi \kappa _{(5)}^2}} {1\over {2(1 +2\xi) ^2}}
(ieB_{C})^2 
-{1\over {1 +2\xi}}\left[
{1\over 8}\left( \nabla_{C}B_{D} -\nabla_{D}B_{C}\right)^2 
+\Lambda_{(5)}
\right]
\Biggr] \cr
&+&{\sigma ^2\over 2}{d^2\over {(d +1)(d -1)}}\times\cr
&&\times
\Biggl[
{ {[-2/d +2(1 +2\xi)](4d/(d -1)) -2}\over {1 +4\xi d/(d -1)}}
{1\over {1/\kappa _{(5)}^2 +\xi \psi ^2[ 1 +4\xi d/(d -1)]}}\cr
&&-{\kappa _{(5)} ^2\over \xi}{{d -1}\over {d +1}}
\left( -{2\over d} +1 +4\xi\right)
\ln\left[ { {1 +\xi \kappa _{(5)} ^2\psi ^2( 1 +2\xi)}\over
{1 +\xi \kappa _{(5)} ^2\psi ^2}[ 1+4\xi d/(d -1)]}\right]\Biggr].
\ea 
We expand the denominator of
the first term in $\sigma,$ 
finding that we can express the effective potential as
\ba
V_{eff}(\psi ^2)
&=&\psi ^2\Biggl\{
{\mu _{(5)} ^2\over 2}{1\over { 1 +2\xi}}
-{\mu _{(5)} ^2\over 4}{1\over { 1 +2\xi}} 
+\lambda _{(5)}{1\over {\xi\kappa _{(5)} ^2}}{1\over {2( 1 +2\xi)^2}}
  \left( {1\over 4} +2\xi\right)\cr
&&-(ieB_{C})^2 {1\over {2( 1 +2\xi)}}
+ie(\nabla_{C}B_{C})\cr
&&
-{1\over 2}\sigma ^2\kappa _{(5)} ^2 (\xi \kappa _{(5)}^2)
 {d^2\over {(d +1)(d -1)}}
  \left[ \left( -{2\over d} +2(1 +2\xi)\right){4d\over {d -1}} -2\right]
\Biggr\}\cr
&+&\psi ^4\Biggl\{ 
{\lambda _{(5)}\over 4} {1\over {1 +2\xi}} 
-{\lambda _{(5)}\over 4}{1\over {4( 1 +2\xi)}}\cr
&&+{1\over 2}\sigma ^2\kappa _{(5)} ^2 {1\over 2}(\xi \kappa _{(5)}^2)^2
 {d^2\over {(d +1)(d -1)}}
  \left[ \left( -{2\over d} +2(1 +2\xi)\right){4d\over {d -1}} -2\right]
   \left( 1 +\xi{4d\over {d -1}}\right)\Biggr\}
\cr
&+&O[\psi ^6]\sigma ^2\kappa _{(5)} ^2 {1\over 6}( \xi \kappa _{(5)}^2)^3\cr
&+&\ln\left[ 1 +\psi ^2\kappa _{(5)} ^2\xi (1 +2\xi)\right]
\times \cr
&&\times
\Biggl[ 
\mu _{(5)} ^2 {1\over {\xi \kappa _{(5)}^2}} {1\over {2(1 +2\xi) ^2}}
 \left( {1\over 2} +2\xi\right)
-\lambda _{(5)} {1\over {( \xi \kappa _{(5)}^2})^2}
 {1\over {2(1 +2\xi)^3}}\left( {1\over 4} +2\xi\right)\cr
&&-{1\over {\xi \kappa _{(5)}^2}} {1\over {2(1 +2\xi) ^2}}
(ieB_{C})^2 
-{1\over {1 +2\xi}}\left[
{1\over 8}\left( \nabla_{C}B_{D} -\nabla_{D}B_{C}\right)^2 
+\Lambda_{(5)}
\right]
\Biggr] \cr
&+&{1\over 2}\sigma ^2\kappa _{(5)} ^2{d^2\over {(d +1)(d -1)}}\times\cr
&&\times
\Biggl[
{ {[-2/d +2(1 +2\xi)](4d/(d -1)) -2}\over {1 +4\xi d/(d -1)}}\cr
&&-{1\over \xi}{{d -1}\over {d +1}}
\left( -{2\over d} +1 +4\xi\right)
\ln\left[ { {1 +\xi \kappa _{(5)} ^2\psi ^2( 1 +2\xi)}\over
{1 +\xi \kappa _{(5)} ^2\psi ^2}[ 1+4\xi d/(d -1)]}\right]\Biggr].
\ea 
At the vev's of the bulk fields, the effective
potential is, up to logarithmic terms, of the form 
$V_{eff} =\mu _{eff} ^2(\psi \psi ^{\ast}/2) 
+\lambda _{eff}(\psi \psi^{\ast})^2 +O[(\psi \psi ^{\ast})^3],$
where
\ba
\mu _{eff} ^2 &\sim&
\mu _{(5)} ^2 
-2\sigma ^2\xi\kappa _{(5)}^4
\\
\lambda _{eff} &\sim&
\lambda_{(5)} +\sigma ^2\xi^2\kappa _{(5)}^6.
\ea

Upon the scalar field taking a non-vanishing vev
$\left<\psi \right>$ (with its conjugate taking the corresponding
conjugate vev $\left<\psi ^{\ast} \right>$)
and the gauge field taking 
$\left<{\bf B}\right> =0,$ with $\nabla_{A} \left<{\bf B}\right> =0,$ 
we find from  Eq.~(\ref{eqn:Einstein-psi})
the same relations for the four-dimensional Planck scale and
the matter fields mass scales as those found in Section \ref{sec:scalar}. 
Consequently, for $\mu_{eff}<0$ and $M_{\psi}\lambda _{eff}>0,$
where $M_{\psi}$ is such that  $\psi =M_{\psi}^{{1\over 2}}\Psi,$
we find that $|\mu_{eff}/(M_{\psi}\lambda _{eff})|$
yields the same constraint relation between $\xi$ 
and $M_{Pl(4)}$ as that found for the case of $\Phi$ in
Eq.~(\ref{eqn:Planck_4}). 
From similar considerations, it follows that the scalar field vev's 
are of order the Planck scale, which implies that both the mass of the
scalar field and the mass of $B_{A}$ generated by the SSB mechanism
are also of order the Planck scale. The mass of $B_{N}$ is of order
$\sigma ^2 \sim (M_{Pl(4)} ^2M_{\phi} ^2)^2.$ Whereas $\Phi$ and $B_A$
become strongly localized about the brane, $B_N$ becomes localized
on the brane by the delta profile of the brane tension.
Notice that, as before, the appearance of a new
mass scale associated with the scalar field is quite natural given
that in the bulk spacetime the self-interaction  
coupling constant, $\lambda_{(5)},$ is not dimensionless.

Moreover, the surviving contributions to the condition for a vanishing
effective cosmological constant are 
\ba
&&\Lambda_{(5)} 
=
-{1\over 2}V(\left<\psi \right>\left<\psi ^{\ast} \right>)
-\xi\left( \left<\psi \right>{\partial V\over \partial \psi} 
+\left<\psi ^{\ast}\right> {\partial V\over \partial \psi ^{\ast}}\right)
 _{\psi =\left<\psi \right>, \psi ^{\ast} =\left<\psi ^{\ast}\right>}
\cr
&-&K^2\Big|_{\psi^2 =\left<\psi \right>^2, 
}
\left[
\left( {1\over \kappa_{(5)}^2} 
+\xi\left<\psi \right>\left< \psi ^{\ast}\right>\right)
 {{d ^2 -d +4}\over {2d ^2}}
+2\xi^2\left<\psi \right>\left< \psi ^{\ast}\right>
 \left( {{-d ^2 +3d +4}\over {2d ^2}} -4\xi\right)\right].\qquad
\ea
Hence, for a vanishing cosmological constant on the brane the   
five-dimensional Planck scale is found to scale like the condition 
in Eq.~(\ref{eqn:Planck_5:Lambda_5}).

\section{Discussion and Conclusions}

In this work we have examined the mechanism of spontaneous symmetry
breaking due to a scalar field in the bulk spacetime coupled non-minimally
to gravity.  
We have shown that a bulk scalar field can be a source 
of symmetry breaking on the brane but only at very high energies. 
We derived the conditions which allow for the existence of a
non-vanishing bulk scalar field vacuum configuration  
and demonstrated that the scales of the induced masses are of order the
four-dimensional Planck scale, thus failing to accommodate on the
brane the typical scales 
of the Standard Model.
We notice, however, that this implies that the bulk scalar fields become very
short range about the position of the brane and thus strongly localized.

In the presence of a bulk gauge field, we found that on
the brane the component $B_A$ interacts with the charged scalar field,
whereas the component $B_N$ interacts solely with the brane tension.
This implies that, upon the acquisition by the scalar field of a
non-vanishing vev, 
the component $B_A$ acquires a mass of order the Planck mass on the
brane and thus becomes 
localized. The component $B_N$ remains massless with respect to 
the interaction
with $\psi,$ acquiring however a ``topological'' mass derived from the
coupling with the delta-localized brane tension.

%


Furthermore, we observe that in the absence of the non-minimal coupling
of the bulk scalar fields to gravity, i.e. for $\xi =0,$  the
effective potential on the brane of a bulk scalar field  reduces in
both cases to  $V_{eff} = \mu _{(5)} ^2( \phi ^2/2) +\lambda _{(5)}( \phi ^4/4)
  V+3\sigma ^2 \kappa _{(5)}^2d^2/[(d +1)(d -1)].$  The realization of
a braneworld universe  as a co-dimension one surface of localized
matter contributes a constant term proportional to the brane tension $\sigma$
to the effective potential.  The brane tension does not, however,
contribute to the mechanism of spontaneous symmetry breaking observed
on the brane unless the bulk scalar fields are non-minimally coupled
to gravity. This is observed in the dependence on the coupling
parameter $\xi$ of the parameters $\mu _{eff}^2$ and $\lambda _{eff}.$
Moreover, the mixing of the discontinuity in  the
extrinsic curvature with the discontinuity in the normal derivative of
the scalar field, as encompassed by the corresponding  matching
conditions, is also  $\xi$--dependent. 
Such mixing is switched off when  $\xi =0,$ as  already
noticed in Ref.~{\cite{BV00}} and also found in Ref.~{\cite{bc06}}.
These observations seem to suggest that the matter localized on the
brane will only interact with bulk matter fields through gravity when
a non-minimal coupling exists. 

On a more technical side, our approach goes a step further in setting up 
the framework for studying the brane induced physics which arises from the 
presence of matter fields in the bulk. The case of a vector field 
coupled non-minimally to gravity was previously considered  
and its implications for Lorentz symmetry on the brane 
studied in Ref. \cite{bc06}. A scalar field coupled non-minimally 
to gravity is treated in the present paper. Our approach allows 
to relate the cosmological constant problem and the scale of gravity
with the mechanism of  
origin of mass, which we have found to relate with the mechanism of
localization of bulk fields on the brane.

\vskip 0.5cm

\centerline{\bf {Acknowledgments}}

\vskip 0.2cm

\noindent 
C. C. thanks Funda\c c\~ao para a Ci\^encia e a 
Tecnologia - FCT (Portuguese Agency) for financial support under the
fellowship /BPD/18236/2004. O. B. acknowledges the partial 
support of the FCT project POCI/FP/63916/2005. The authors would like to thank 
Martin Bucher, Mariam Bouhmadi Lopez, Fernando Quevedo and Kyriakos
Tamvakis for useful discussions. 



\end{document}